\documentclass[11pt]{article}
\usepackage{amssymb}

\let\ensuremathTEMP\ensuremath

\makeatletter%
\def\nottoobig#1{{\hbox{$\left#1\vcenter to1.111\ht\strutbox{}\right.\n@space$}}}
\makeatother%

\makeatother%

\makeatletter%

\newcount\hour  \newcount\minutes  \hour=\time  \divide\hour by 60
\minutes=\hour  \multiply\minutes by -60  \advance\minutes by \time
\def\mmmddyyyy{\ifcase\month\or Jan\or Feb\or Mar\or Apr\or May\or Jun\or Jul\or
  Aug\or Sep\or Oct\or Nov\or Dec\fi \space\number\day, \number\year}
\def\hhmm{\ifnum\hour<10 0\fi\number\hour :%
  \ifnum\minutes<10 0\fi\number\minutes}
\def\Draft{{\it Draft of \mmmddyyyy}}

\def\ps@jtsheadings{%
\def\@oddhead{\it\rightmark\hfil\rm\thepage}%
\def\@oddfoot{\hfil\Draft}%
\if@twoside%
\def\@evenhead{\rm\thepage\hfil\it\leftmark}%
\def\@evenfoot{\Draft\hfil}%
\else
\let\@evenhead\@oddhead%
\let\@evenfoot\@oddfoot%
\fi%
}
\def\ps@jtsplain{%
\def\@oddhead{\hfil\Draft}%
\def\@oddfoot{\hfil\rm\thepage\hfil}%
\let\@evenfoot\@oddfoot%
\if@twoside \def\@evenhead{\Draft\hfil} \else \let\@evenhead\@oddhead \fi
}

\def\chaptermark#1{\markboth{\thechapter.\ #1}{\thechapter.\ #1}}%
\def\sectionmark#1{\markright{\thesection.\ #1}}

\def\section{\@startsection {section}{1}{\z@}
    {3.5ex plus1ex minus.2ex}{2.3ex plus.2ex}{\Large\bf}}
\def\subsection{\@startsection{subsection}{2}{\z@}
    {3.25ex plus1ex minus.2ex}{1.5ex plus.2ex}{\large\bf}}
\def\subsubsection{\@startsection{subsubsection}{3}{\z@}
    {3.25ex plus1ex minus.2ex}{1.5ex plus.2ex}{\normalsize\bf}}
\def\paragraph{\@startsection{paragraph}{4}{\z@}
    {3.25ex plus1ex minus.2ex}{1em}{\normalsize\bf}}
\def\subparagraph{\@startsection{subparagraph}{4}{\parindent}
    {3.25ex plus1ex minus.2ex}{1em}{\normalsize\bf}}

\makeatother%

\makeatletter \@beginparpenalty=10000 \makeatother

\def\underl#1 {\leavevmode\let\first=\relax\underli #1 }
\def\underli#1 {\ifx&#1\let\next=\relax\unskip
                \else\let\next=\underli\first\ulinebox{#1}\fi\let\first=\undersp\next}
\def\undersp{\penalty50\ulinebox{\space}\penalty50}
\def\ulinebox#1{\vtop{\hbox{\strut#1}\hrule}}%
\def\unice#1 {\underl #1 & }
\def\desclabel#1{\bf #1\hfil}
\def\desc{\list{}{%
\labelwidth= \leftmargin
\advance \labelwidth by -\labelsep
\let \makelabel=\desclabel}}

\makeatletter %

\newlength{\filength}
\newsavebox{\gcbox}
\sbox{\gcbox}{\framebox[\filength]{\rule{0ex}{2ex}}}

\newlength{\leftjustindent}
\newlength{\@leftjustindent}
\def\leftjust{\let\\\@leftjustcr\let\end\@endleftjust
  \addtolength{\@leftjustindent}{\leftjustindent} \vcenter\bgroup
\halign\bgroup \hbox to\displaywidth{
\rule{\@leftjustindent}{0ex}$\displaystyle##$\hfill }\crcr }
\def\endleftjust{\crcr\egroup\egroup\endgroup}
\def\@endleftjust#1{\crcr\egroup\egroup\@checkend{#1}\endgroup}
\def\@leftjustcr{\crcr}

\newtheorem{theorem}{Theorem}[section]
\newtheorem{corollary}[theorem]{Corollary}

\newcommand{\qedblob}{\mbox{\rule[-1.5pt]{5pt}{10.5pt}}}
\def\literalqed{{\ \nolinebreak\hfill\mbox{\qedblob\quad}}}

\def\qed{\literalqed}

\newtheorem{lemma}[theorem]{Lemma}

\newcommand{\singlespacing}{\let\CS=
\@currsize\renewcommand{\baselinestretch}{1}\tiny\CS}
\newcommand{\singlespacingplus}{\let\CS=
\@currsize\renewcommand{\baselinestretch}{1.25}\tiny\CS}
\newcommand{\doublespacing}{\let\CS=
\@currsize\renewcommand{\baselinestretch}{1.75}\tiny\CS}
\newcommand{\draftspacing}{\let\CS=
\@currsize\renewcommand{\baselinestretch}{2.0}\tiny\CS}

\makeatother%

\mathcode`\0="0030      %
\mathcode`\1="0031
\mathcode`\2="0032
\mathcode`\3="0033
\mathcode`\4="0034
\mathcode`\5="0035
\mathcode`\6="0036
\mathcode`\7="0037
\mathcode`\8="0038
\mathcode`\9="0039
\singlespacingplus

\newtheorem{definition}[theorem]{Definition}

\makeatletter
\clubpenalty=\@highpenalty
\widowpenalty=\@highpenalty
\makeatother

\makeatletter
\newcommand{\niceonespacing}{\let\CS=\@currsize\renewcommand{\baselinestretch}{1.1}\tiny\CS}
\newcommand{\nicezeroonespacing}{\let\CS=\@currsize\renewcommand{\baselinestretch}{1.05}\tiny\CS}
\newcommand{\nicetwospacing}{\let\CS=\@currsize\renewcommand{\baselinestretch}{1.2}\tiny\CS}
\newcommand{\nicethreespacing}{\let\CS=\@currsize\renewcommand{\baselinestretch}{1.3}\tiny\CS}
\newcommand{\singlespacingplusplus}{\let\CS=\@currsize\renewcommand{\baselinestretch}{1.35}\tiny\CS}
\newcommand{\nicefourspacing}{\let\CS=\@currsize\renewcommand{\baselinestretch}{1.4}\tiny\CS}
\newcommand{\nicefivespacing}{\let\CS=\@currsize\renewcommand{\baselinestretch}{1.5}\tiny\CS}
\newcommand{\nicesixpacing}{\let\CS=\@currsize\renewcommand{\baselinestretch}{1.6}\tiny\CS}
\makeatother

\makeatletter
\def\@cite#1#2{[#1\if@tempswa , #2\fi]}
\makeatother

\makeatletter
\def\@citex[#1]#2{\if@filesw\immediate\write\@auxout{\string\citation{#2}}\fi
  \def\@citea{}\@cite{\@for\@citeb:=#2\do
    {\@citea\def\@citea{,\linebreak[0]}\@ifundefined
       {b@\@citeb}{{\bf ?}\@warning
       {Citation `\@citeb' on page \thepage \space undefined}}%
\hbox{\csname b@\@citeb\endcsname}}}{#1}}
\makeatother

\makeatletter
\def\ps@thesis{\def\@oddhead{\hfil\rm\thepage\hfil}\def\@oddfoot{}\def\@evenhead{\hfil\rm\thepage\hfil}\def\@evenfoot{}\def\chaptermark##1{}\def\sectionmark##1{}}
\makeatother

\makeatletter
\def\foobarpt{\textfont\z@\tenrm 
  \scriptfont\z@\ninrm \scriptscriptfont\z@\sevrm
\textfont\@ne\tenmi \scriptfont\@ne\ninmi \scriptscriptfont\@ne\sevmi
\textfont\tw@\tensy \scriptfont\tw@\ninsy \scriptscriptfont\tw@\sevsy
\textfont\thr@@\tenex \scriptfont\thr@@\tenex \scriptscriptfont\thr@@\tenex
\def\unboldmath{\everymath{}\everydisplay{}\@nomath\unboldmath
          \textfont\@ne\tenmi 
          \textfont\tw@\tensy \textfont\lyfam\tenly
          \@boldfalse}\@boldfalse
\def\boldmath{\@ifundefined{tenmib}{\global\font\tenmib\@mbi\@magscale1\global
        \font\tensyb\@mbsy \@magscale1\global\font
         \tenlyb\@lasyb\@magscale1\relax\@addfontinfo\@xiipt
              {\def\boldmath{\everymath
                {\mit}\everydisplay{\mit}\@prtct\@nomathbold
                \textfont\@ne\tenmib \textfont\tw@\tensyb 
                \textfont\lyfam\tenlyb\@prtct\@boldtrue}}}{}\@xiipt\boldmath}%
\def\prm{\fam\z@\tenrm}%
\def\pit{\fam\itfam\tenit}\textfont\itfam\tenit \scriptfont\itfam\ninit
   \scriptscriptfont\itfam\sevit
\def\psl{\fam\slfam\tensl}\textfont\slfam\tensl 
     \scriptfont\slfam\tensl \scriptscriptfont\slfam\tensl
\def\pbf{\fam\bffam\tenbf}\textfont\bffam\tenbf 
   \scriptfont\bffam\ninbf \scriptscriptfont\bffam\ninbf 
\def\ptt{\fam\ttfam\tentt}\textfont\ttfam\tentt
   \scriptfont\ttfam\nintt \scriptscriptfont\ttfam\nintt 
\def\psf{\fam\sffam\tensf}\textfont\sffam\tensf
    \scriptfont\sffam\tensf \scriptscriptfont\sffam\tensf
\def\psc{\@getfont\psc\scfam\@xiipt{\@mcsc\@magscale1}}%
\def\ly{\fam\lyfam\tenly}\textfont\lyfam\tenly 
   \scriptfont\lyfam\ninly \scriptscriptfont\lyfam\sevly
 \@setstrut \rm}

\makeatother

\newcommand{\naturalnumber}{\ensuremath{{  \mathbb{N} }}}

\newcommand{\sat}{{\rm SAT}}

\newcommand{\up}{{\rm UP}}

\newcommand{\fewp}{{\rm FewP}}
\newcommand{\coup}{{\rm coUP}}

\newcommand{\p}{{\rm P}}
\newcommand{\littlep}{{p}}

\newcommand{\np}{{\rm NP}}

\newcommand{\conp}{{\rm coNP}}

\newcommand{\sigmak}{\ensuremath{\Sigma_k^p}}

\singlespacingplus

\newcommand{\protm}{\ensuremath{{\mbox{\it{}Protectcodings}_m}}}
\newcommand{\canmac}{\ensuremath{{\mbox{\it{}CAN}}}}
\newcommand{\manyonea}{\ensuremath{\,\leq_{m}^{{\littlep},\,A}\,}}
\newcommand{\manyone}{\ensuremath{\,\leq_{m}^{{\littlep}}\,}}
\newcommand{\manyonered}{\ensuremath{\,\leq_{m}^{{\littlep}}\,}}

\newcommand{\snonettred}{\ensuremath{\,\leq_{1\hbox{-}{tt}}^{\cal SN}\,}}
\newcommand{\snaonettred}{\ensuremath{\,\leq_{1\hbox{-}{tt}}^{{\cal SN},A}\,}}

\newcommand{\onettred}{\ensuremath{\,\leq_{1\hbox{-}{tt}}^{{\littlep}}\,}}
\newcommand{\onettreda}{\ensuremath{\,\leq_{1\hbox{-}{tt}}^{{\littlep, A}}\,}}

\newcommand{\twottreda}{\ensuremath{\,\leq_{2\hbox{-}{tt}}^{{\littlep},A}\,}}

\newcommand{\sigmastar}{\ensuremath{\Sigma^\ast}}

\newcommand{\calc}{\ensuremath{{\cal C}}}
\newcommand{\cald}{\ensuremath{{\cal D}}}
\newcommand{\calca}{\ensuremath{{{\cal C}^A}}}
\newcommand{\npa}{\ensuremath{\np^A}}

\newcommand{\fpta}{\ensuremath{\fp_{\rm t}^A}}
\newcommand{\fpt}{\ensuremath{\fp_{\rm t}}}

\newcommand{\conpa}{\ensuremath{\conp^A}}

\newcommand{\condition}{\, \mid \:}

\singlespacingplus
\makeatletter
\def\@listI{\leftmargin\leftmargini \parsep 4.5pt plus 1pt minus 1pt\topsep
6pt plus 2pt minus 2pt \itemsep  2pt plus 2pt minus 1pt}

\let\@listi\@listI
\@listi
\makeatother

\newcommand{\npsvt}{\ensuremath{{\rm NPSV}_{\rm t}}}
\newcommand{\npsvta}{\ensuremath{{\rm NPSV}^A_{\rm t}}}

\makeatletter %
 \newcommand{\setoffdisplay}{\rule{5.9in}{1pt}}

\makeatother



\newcommand{\fp}{\ensuremathTEMP{{\rm FP}}}

\newcommand{\sigmai}{\ensuremathTEMP{{\Sigma_{i}^p}}}

\newcommand{\sigmaj}{\ensuremathTEMP{{\Sigma_j^p}}}

\newcommand{\pik}{\ensuremathTEMP{{\Pi_k^p}}}
\newcommand{\pcd}{\ensuremathTEMP{\rm P}^{{\cal C} : {\cal D}}}

\newcommand{\pconectwo}{\ensuremathTEMP{\rm P}^{{\cal C}_1 : {\cal C}_2}}
\newcommand{\pctwocone}{\ensuremathTEMP{\rm P}^{{\cal C}_2 : {\cal C}_1}}

\newcommand{\pnpone}{\ensuremathTEMP{\p^{\np[1]}}}

\newcommand{\funny}{\ensuremathTEMP{\p^{\np \cap \conp :\np}}}
\newcommand{\funnycalc}{\ensuremathTEMP{\p^{\np \cap \conp :{\cal C}}}}
\newcommand{\funnyrev}{\ensuremathTEMP{\p^{\np : \np \cap \conp}}}
\newcommand{\funnytt}{\ensuremathTEMP{\p^{(\np \cap \conp ,\np)}}}
\newcommand{\partialfuna}{\ensuremathTEMP{\p^{(\npa \cap \conpa ,\npa)}}}
\newcommand{\funnyttcalc}{\ensuremathTEMP{\p^{(\np \cap \conp ,{\cal C})}}}
\newcommand{\funnytta}{\ensuremathTEMP{\left(\p^{(\np \cap \conp ,\np)}\right)^A}}
\newcommand{\smartfunny}{\ensuremathTEMP{\p^{\{\np \cap \conp\} :\np}}}
\newcommand{\smartfunnyrev}{\ensuremathTEMP{\p^{\np : \{\np \cap \conp\}}}}
\newcommand{\smartfunnytt}{\ensuremathTEMP{\p^{(\{\np \cap \conp\} ,\np)}}}

\newcommand{\fixupa}{\ensuremathTEMP{{ \mbox{} \cup B'}}}
\newcommand{\fixupz}{\ensuremathTEMP{{ \mbox{} \cup B''}}}
\newcommand{\rsnnp}{\ensuremathTEMP{{\rm R}_{1\hbox{-}{tt}}^{\cal SN}(\np)}}
\newcommand{\rsncalc}{\ensuremathTEMP{{\rm R}_{1\hbox{-}{tt}}^{\cal SN}({\cal C})}}
\newcommand{\rsnnpa}{\ensuremathTEMP{\left({\rm R}_{1\hbox{-}{tt}}^{\cal SN}(\np)\right)^A}}
\newcommand{\rspecialtwoa}{\ensuremathTEMP{\left({\rm R}_{2\hbox{-}{tt}}^{p}(\np)\right)^A}}
\newcommand{\rponettnp}{\ensuremathTEMP{{\rm R}_{1\hbox{-}{tt}}^p(\np)}}
\newcommand{\rptwottnp}{\ensuremathTEMP{{\rm R}_{2\hbox{-}{tt}}^p(\np)}}
\newcommand{\quadruple}{\ensuremathTEMP{{\langle i,j,k,l \rangle}}}
\newcommand{\quadruplehats}{\ensuremathTEMP{{\langle \widehat{i},
        \widehat{j},\widehat{k},\widehat{l} \rangle}}}

\newcommand{\npinterconp}{\ensuremathTEMP{{\rm NP} \cap {\rm coNP}}}
\newcommand{\dpintercodp}{\ensuremathTEMP{{\rm DP} \cap {\rm coDP}}}

\newcommand{\proof}{\noindent {\bf Proof:}\quad}

\renewcommand{\land}{\,\wedge\,}
\renewcommand{\lor}{\,\vee\,}

\begin{document}

\typeout{WARNING:  BADNESS used to suppress reporting.  Beware.}
\hbadness=3000%
\vbadness=10000 %

\pagestyle{empty}
\pagestyle{empty}
\setcounter{footnote}{0}

\title{\protect\rsnnp\ Distinguishes Robust Many-One 
and Turing Completeness}

\author{%
Edith Hemaspaandra\thanks{\protect\singlespacing{}Supported in part 
by grant
NSF-INT-9513368/DAAD-315-PRO-fo-ab.  Work done in part while 
visiting 
Friedrich-Schiller-Universit\"at Jena.
Email: {\tt edith@bamboo.lemoyne.edu}.}
\\Department of Mathematics\\
Le Moyne College\\
Syracuse, NY 13214, USA
\and 
Lane A. Hemaspaandra\thanks{\protect\singlespacing{}Supported in part 
by grants NSF-CCR-9322513 and 
NSF-INT-9513368/DAAD-315-PRO-fo-ab.  Work done in part while 
visiting 
Friedrich-Schiller-Universit\"at Jena.
Email: {\tt lane@cs.rochester.edu}.}
\\Department of Computer Science\\University of Rochester\\
            Rochester, NY 14627, USA
\and 
Harald Hempel\thanks{\protect\singlespacing{}Supported in part 
by grant
NSF-INT-9513368/DAAD-315-PRO-fo-ab.
Work done in part while 
visiting 
Le~Moyne College.
Email: {\tt hempel@\protect\linebreak[0]informatik.\protect\linebreak[0]uni-jena.de}.}
\\Institut f\"ur Informatik\\
Friedrich-Schiller-Universit\"at Jena\\
07740 Jena, Germany}

\date{September, 1996; Revised July 15, 1997}

{

\singlespacing

\maketitle

}

{\singlespacing

\begin{abstract}
Do complexity classes have many-one complete sets if and only if
they have Turing-complete sets?  We prove that there is a relativized
world in which a relatively natural complexity class---namely a
downward closure of NP, \rsnnp---has Turing-complete sets but has no
many-one complete sets.  In fact, we show that in the same relativized
world this class has 2-truth-table complete sets but lacks
1-truth-table complete sets.  As part of the groundwork for our result, 
we prove that \rsnnp\ has many equivalent forms having to do
with ordered and parallel access to $\np$ and $\npinterconp$.
\end{abstract}

} %

\niceonespacing
\setcounter{page}{1}
\pagestyle{plain}
\sloppy

\section{Introduction}

In this paper, we ask whether there are natural complexity classes for
which the {\em existence\/} of many-one and Turing-complete sets can
be distinguished.  Many standard complexity classes---e.g., R, BPP, UP,
FewP, $\npinterconp$---are known that in some relativized worlds lack
many-one complete (m-complete) sets, and that in some relativized
worlds lack Turing-complete (T-complete) sets.  However, for none of
the classes just mentioned is there known any relativized world in
which the class (simultaneously) has T-complete sets but lacks
m-complete sets.  In fact, for $\npinterconp$ and BPP,
Gurevich~\cite{gur:c:comp} and 
Ambos-Spies~\cite{amb:j:complete-problems}
respectively have shown that no such world can exist.  In this paper, we
will show that there is a downward closure of NP, \rsnnp, that
in some relativized worlds simultaneously has T-complete sets and
lacks m-complete sets.

In fact, $\rsnnp$ has even stronger properties.  We will see that it
{\em robustly\/}---i.e., in all relativized worlds, including the real
world---has 2-truth-table complete 
(2-tt-complete) sets.  Yet we will see that in our
relativized world it lacks 1-tt-complete sets.  Thus, 
this class displays a very crisp borderline
between those reduction types under which it robustly has complete
sets, and those reduction types under which it does not robustly have
complete sets.

We now turn in more detail to describing what is currently 
known in the literature regarding robust completeness.
Sipser~\cite{sip:c:complete-sets} first studied this notion, and showed 
that $\np \cap \conp$ and random polynomial time (R) do not 
robustly have m-complete 
sets.
However, as alluded to in the first paragraph,
Gurevich~\cite{gur:c:comp} proved that, in each relativized 
world, $\np \cap \conp$ has m-complete sets if and only if 
$\np \cap \conp$ has T-complete sets.
Thus, $\np \cap \conp$ cannot distinguish robust m-completeness from 
robust T-completeness. 
Ambos-Spies~\cite{amb:j:complete-problems} extended this by 
showing that no class closed downwards under Turing reductions can 
distinguish robust m-completeness from robust T-completeness.

Thus, the only candidates for distinguishing robust m-completeness from 
robust T-completeness within PSPACE are those classes in PSPACE that may 
lack m-complete sets yet that seem not to be closed downwards 
under 
Turing reductions.
The classes R, \up, and \fewp\ have been shown to 
potentially be of this form 
(see, respectively, \cite{sip:c:complete-sets}, \cite{har-hem:j:up}, 
and \cite{hem-jai-ver:j:up-turing} for proofs that these classes do not 
robustly have 
m-complete sets\footnote{\protect\singlespacing{}The study 
of robust completeness
has been 
pursued in many papers.  Of particular 
interest is the elegant work of 
Bovet, Crescenzi, and Silvestri~\protect\cite{bov-cre-sil:j:uniform},
which abstracts the issue of m-completeness away from particular
classes via general conditions.
Also, the other method of proving such results has been 
reasserted, in a very abstract and algebraic form, in the recent
thesis of Borchert~\protect\cite{bor:thesis:promise},
which re-poses abstractly the proof approach that was pioneered
by Sipser~(\protect\cite{sip:c:complete-sets},
see also~\protect\cite{reg:unpub:cons}).  
Like the Bovet/Crescenzi/Silvestri approach, this method 
abstracts away from directly addressing completeness, 
in the case of this approach via characterizing 
completeness in terms of the issue of the existence of 
certain index sets (in the Borchert version, the discussion 
is abstracted one level further than this).
In Section~\ref{s:completeness} we 
follow the 
Sipser/Regan/Borchert ``index sets/enumeration'' approach, 
in its non-algebraic formulation.}).
Unfortunately, these classes are also known to not robustly have 
T-complete sets~\cite{hem-jai-ver:j:up-turing}, and 
so 
these classes fail to distinguish robust m-completeness from 
robust T-completeness.

In fact, to the best of our knowledge, the literature contains only one 
type of class that distinguishes robust m-completeness from robust 
T-completeness---and that type is deeply unsatisfying. The type is 
certain ``union'' classes---namely, certain classes that either union 
incomparable classes or that union certain infinite hierarchies of 
bounded-access classes. 
Both exploit the fact that if such classes have some m-complete set it 
must fall into some particular element of the union. An example 
of the ``incomparable'' case is that if $\np \cup \conp$ has m-complete 
sets then $\np=\conp$ (and $\np = \conp$ is not robustly 
true~\cite{bak-gil-sol:j:rel}).
An example (from~\cite{hem-jai-ver:j:up-turing}) of the ``infinite union 
of bounded-access classes'' case is the boolean 
hierarchy~\cite{cai-gun-har-hem-sew-wag-wec:j:bh1}, i.e., 
$${\rm BH}=\{L \condition L \leq^p_{{btt}} 
\sat\}.$$
From its definition, it is clear that \sat\ is T-complete 
(indeed, even bounded-truth-table complete) for ${\rm BH}$\@.  However,
if BH had an m-complete set then that set 
(since it would be in BH) would have to be computable 
via some $k$-truth-table reduction to \sat, so there would be a $\hat{k}$ 
such that ${\rm BH}=\{L \condition L \leq^p_{\hat{k}\hbox{-}{tt}} 
\sat\}$, but this is known to not be robustly 
true~\cite{cai-gun-har-hem-sew-wag-wec:j:bh1}.

We at this point mention an interesting related topic that this paper
is not about, and with
which our work should not
be confused.  That topic, in contrast to our attempt to distinguish
{\em the existence of\/} m-complete and T-complete sets for a class,
is the study of whether one can merely distinguish {\em the set of\/}
m-complete and T-complete sets for a class.  For example, 
various conditions 
(most strikingly, NP does not ``have 
p-measure~0''~\cite{lut-may:j:cook-karp}) 
are known such that their truth would imply
that the class of NP-m-complete sets
differs from the class of NP-T-complete sets.  
However, this does not
answer our question, as NP robustly has m-complete sets and robustly
has T-complete sets.  The exact same comment applies to the work of
Watanabe and Tang~\cite{tan-wat:j:pspace} that shows certain
conditions under which the class of PSPACE-m-complete sets differs
from the class of PSPACE-T-complete sets.  Also of interest, but not
directly related to our interest in the {\em existence\/} of complete
sets, is the work of Longpr\'e and Young~\cite{lon-you:j:fast} showing
that within \np\ Turing
reductions can be polynomially ``faster'' than many-one reductions.

As mentioned at the start of this section, in this paper we prove that
\rsnnp\ robustly has T-complete sets but does not robustly have
m-complete sets.  We actually prove the stronger result
that \rsnnp\ distinguishes robust 1-tt-completeness from robust
2-tt-completeness.  This of course implies that there is 
a relativized world in which $\rsnnp$ has T-complete 
(even 2-tt-complete) sets but lacks m-complete (even 1-tt-complete)
sets.
It is important to note that this is not
analogous to the ``union'' examples given two paragraphs ago.
$\rsnnp$ is not a ``union'' class.  Also, the mere fact that a class is
defined in terms of some type of access to \np\ is not, in and of
itself, enough to preclude robust m-completeness, as should be clear
from the fact that $\rponettnp$ and $\rptwottnp$ 
robustly have m-complete sets
(note: $\rponettnp
\subseteq \rsnnp \subseteq \rptwottnp$).

Regarding the background of the reducibility \snonettred, we mention
that Homer and
Longpr\'e (\cite[Corollary~5]{hom-lon:j:sparse}, see 
also~\cite{ogi-wat:j:pbt})
have recently proven that if any set
that is
\manyonered-hard for \np\ is \snonettred-reducible (or even
$\leq_{{btt}}^{\cal SN}$-reducible) to a sparse set then the
polynomial hierarchy equals \np\@.  Regarding the class \rsnnp, we
consider \rsnnp\ to be its most natural form.  However,
Section~\ref{s:lots} proves that this class has many equivalent
characterizations (for example, it is exactly the class
\funnytt---what a
\p\ machine can compute via one \npinterconp\ query made in parallel with 
one \np\ query).  Section~\ref{s:lots}
also gives a candidate language for \rsnnp\ 
(namely PrimeSAT
$=\{\langle i,F \rangle \condition i \in {\rm PRIMES} \iff F \in \sat\}$)
and notes that though $\rsnnp \subseteq {\rm DP}$,\footnote{${\rm DP} 
=\{L \condition (\exists L_1, L_2 \in 
\np)[L=L_1-L_2]\}$~\cite{pap-yan:j:dp}.}
the containment is strict unless the polynomial hierarchy collapses.

\section{Preliminaries}\label{s:prelim}

For standard notions not defined here, we refer
the reader to any computational complexity
textbook, 
e.g.,~\cite{bov-cre:b:complexity,pap:b:complexity,bal-dia-gab:b:sctI-2nd-ed}.

Unless otherwise stated or otherwise obvious from context,
all strings
will use the alphabet
$\Sigma = \{0,1\}$ and all sets will be 
collections of such strings.  For every set $A$ 
we will denote the characteristic function 
of $A$ by $\chi_A$.  $A^{\leq k}$ denotes 
$\{ x \condition x \in A \land |x| \leq k\}$.
Strong nondeterministic reductions were introduced by
Selman~\cite{sel:j:enumeration-reducibility} 
(with different nomenclature)
and
Long~\cite{lon:j:nondeterministic-reducibilities}.
The literature contains two potentially different 
notions of strong nondeterministic {\em truth-table\/}
reducibility, one due to 
Long~\cite{lon:j:nondeterministic-reducibilities}
and the other due to 
Rich~\cite{ric:j:positive} and Homer and 
Longpr\'{e}~\cite{hom-lon:j:sparse}.
(The
notions differ, for example, regarding whether the query generation
is single-valued or multivalued.)
Throughout this paper, we use the notion of Homer and Longpr\'{e}
and Rich.

\begin{definition}\label{d:npsvt} 
(see~\cite{sel:j:taxonomy,sel-mei-boo:j:pos})\ 
A function $f$ is in $\npsvt$ if there exists a 
nondeterministic polynomial-time Turing machine $N$
such that, on each input $x$, it holds that 
\begin{enumerate}
\item at least one computation path of $N(x)$ is an 
accepting path that outputs $f(x)$, and
\item every accepting computation path of $N(x)$ computes
the same value, i.e., $f(x)$.  (Note:  rejecting computation 
paths are viewed as having no output.)
\end{enumerate}
\end{definition}

\begin{definition}\label{d:sn}
(\cite{hom-lon:j:sparse}, see also \cite{ric:j:positive})\ For any 
constant $k$ we say $A$ is $k$-truth-table strong 
nondeterministic reducible to $B$ 
($A \leq_{k\hbox{-}{tt}}^{\cal SN} B$) 
if there is a function in $\npsvt$ that computes both (a) $k$ strings 
$x_1,x_2,\cdots,x_k$ and (b) a predicate, $\alpha$, of $k$ boolean 
variables, such that $x_1,x_2,\cdots,x_k$ and $\alpha$ satisfy:
$$x \in A \iff \alpha(\chi_B(x_1),\chi_B(x_2),\cdots ,\chi_B(x_k)).$$
\end{definition}

Let ${\cal C}$ be a complexity class. We say $A \leq_m^{p,{\cal C}[1]} B$
if and only if there is a function $f \in {\rm FP}^{{\cal C}[1]}$ 
(i.e.,  computable via a deterministic polynomial-time 
Turing machine allowed one query to some oracle
from $\cal C$) such that,
for all $x$, $x\in A \iff f(x) \in B$.

As is standard in the literature, for 
any strings of symbols $a$ and $b$ for which $\leq_a^b$
is defined and any class $\cal C$, let $ {\rm R}_a^b({\cal C}) =
\{L\condition (\exists C \in {\cal C})[L\leq_a^b C]\}$.

Let $\langle \cdot,\cdot \rangle$ be any 
fixed pairing function with the 
standard nice properties (polynomial-time computability, 
polynomial-time invertibility).

We use DPTM (NPTM) as shorthand for
``deterministic (nondeterministic) polynomial-time 
oracle Turing machine,'' and we treat non-oracle Turing
machines as oracle Turing machines that merely happen not 
to use their oracle tapes.
Without loss of generality, 
we henceforward assume that DPTMs and NPTMs are clocked with clocks that are
independent of the oracle.
$M^A(x)$ denotes the computation of the DPTM $M$ with oracle $A$ on input $x$.
At times, when the oracle is clear from context, we may 
write $M(x)$, omitting the oracle superscript(s) (such as $M^A(x)$).

Let $\{M_i\}$ and $\{N_i\}$ respectively be enumerations of deterministic and 
nondeterministic polynomial-time oracle 
Turing 
machines.
Without loss of generality, let these enumerations 
be such that $M_i$ and $N_i$ run in (respectively, deterministic 
and nondeterministic) time $n^i+i$ and let them also be
such that given
$i$ one can in polynomial time 
derive (as Turing machine code) $M_i$ and $N_i$.
 
\begin{definition} \label{d:new}
Let ${\cal C}$ and ${\cal D}$ be complexity classes. 
\begin{enumerate}
\item~\cite{hem-hem-wec:jtoappear:query-order-bh} 
Let $M^{A:B}$ denote a DPTM $M$ making one query to oracle $A$ followed 
by one query to oracle $B$.\footnote{\protect\singlespacing{}We do 
not describe 
the mechanics of having two oracles, as any natural approach
will do in the contexts with which we 
are dealing.  For example, oracle machines can all have one oracle
tape, with the query to the first oracle being contained in the 
tape cells to the right of the origin and the query to the second 
oracle being contained in the tape cells to the left of the 
origin, and with only the appropriate half being erased after
entering the distinguished state denoting a query to that half.
Alternatively and perhaps more naturally, one can allow the 
oracle machine to have one oracle tape per oracle.}
Let
\begin{eqnarray*}
\pcd &=&
\{L \condition (\exists C \in {\cal C})(\exists D \in 
{\cal D})(\exists \ {\rm DPTM }~M)[L=L\left( M^{C : D}\right) ]\} 
\end{eqnarray*}
\item~\cite{hem-hem-hem:ctoappear:query-order-ph} \label{p:tt-order}
Let $M^{(A,B)}$ denote a DPTM $M$ making,
simultaneously, one query to oracle $A$ and one query to oracle $B$.
Let
$${\p^{({\cal C},{\cal D})}}=
\{L  \condition (\exists C \in {\cal C})(\exists D \in 
{\cal D})(\exists \ {\rm DPTM}~M)[L=L\left( M^{(C,D)}\right) ]\}.$$
\end{enumerate}
\end{definition}

Classes of the form $\pcd$ 
were introduced and studied
by Hemaspaandra, Hempel,
and Wechsung~\cite{hem-hem-wec:jtoappear:query-order-bh}.
They focused on the case in which
$\cal C$ and $\cal D$ are levels of the boolean hierarchy.
The present 
authors~\cite{hem-hem-hem:ctoappear:query-order-ph} 
first studied the case in which
$\cal C$ and $\cal D$ are levels of the polynomial hierarchy.
These papers
propose and study the effect of the {\em order\/} of 
database
access on the power of database-accessing
machines.
That line of research has led recently to the counterintuitive
downward collapse result that, 
for each $k \geq 2$, $ \sigmak = \pik \iff \p^{\sigmak[1]} =
\p^{\sigmak[2]}$
(\cite{hem-hem-hem:jtoappear:downward-translation,buh-for:t:two-queries},
see also \cite{hem-hem-hem:t:translating-downwards}), and to
a number of other interesting 
results~\cite{wag:t:parallel-difference,bei-cha:ctoappear:commutative-queries}.

Part~\ref{p:tt-order} 
of Definition~\ref{d:new} is somewhat related to work of
Selivanov~\cite{sel:c:refined-PH}.  
This fact, and the comments of the rest of this
paragraph, were noted
independently by an 
earlier version of the present 
paper~\cite{hem-hem-hem:cConferenceCite:sn-1tt-np-completeness}
and by 
Klaus Wagner
(\cite{wag:t:parallel-difference}, see 
also~\cite{bei-cha:ctoappear:commutative-queries}), whose
observations are in a more general
form (namely, applying to more than two sets and to more 
abstract classes).
We now discuss
the basic facts 
known
about the relationship between 
the classes of Selivanov (for the case of 
``$\triangle$''s of two sets;  see Wagner~\cite{wag:t:parallel-difference} 
for the case of more than two sets) and the classes discussed in
this paper.
Selivanov
studied refinements of the
polynomial hierarchy.
Among the classes he considered, those closest 
to the classes we study in this paper are his 
classes
$$ \sigmai \mbox{\boldmath$\bf \triangle$} \sigmaj
=
\{ L \condition 
(\exists A \in \sigmai)
(\exists B \in \sigmaj)
[ L = A \triangle B]\},$$
where $A \triangle B = (A - B) \cup (B - A)$.
Note, however, that his classes seem to be
different from our classes.  
This can be immediately seen from the fact that all our 
classes are closed under complementation, but the
main theorem of Selivanov
(\cite{sel:c:refined-PH}, see also the 
discussion and
strengthening in~\cite{hem-hem-hem:t:translating-downwards})
states that  no 
class of the form 
$\sigmai 
\mbox{\boldmath$\bf \triangle$} \sigmaj$,
with $i>0$ and $j>0$, 
is closed under complementation unless the 
polynomial hierarchy collapses.  Nonetheless,
the class
$\sigmai 
\mbox{\boldmath$\bf \triangle$} \sigmaj$
is not too much weaker than
$\p^{(\sigmai,\sigmaj)}$,
as it is 
not hard to see (by 
easy manipulations 
if 
$i \neq j$, and from the work of 
Wagner~\cite{wag:j:bounded} and K\"obler,
Sch\"oning, and Wagner~\cite{koe-sch-wag:j:diff}
for the $i=j$ case)
that, for all $i$
and $j$, it holds that $\{ L \condition (\exists L' \in
\sigmai 
\mbox{\boldmath$\bf \triangle$} \sigmaj) [L
\leq_{1\hbox{\rm{}-}\rm{}tt}^p L']\} = 
\p^{(\sigmai,\sigmaj)}$.

\section{Equivalent forms of 
\protect\boldmath{${\bf R}^{\cal SN}_{1\hbox{-}tt}({\bf NP})$}}\label{s:lots}

In this section, we consider the class \rsnnp\ and note that this
class is quite oblivious to definitional variations; it has many
equivalent forms.

The following lemma is from~\cite{hem-hem-hem:ctoappear:query-order-ph} 
and will be useful
here.\footnote{\protect\singlespacing{}The theorem, its 
asymmetry notwithstanding,
is not mistyped.  One does not 
need to additionally assume that
${\cal C}_2$ is closed downwards under 
$\leq_m^{p,{\cal C}_1[1]}$.}

\begin{lemma} 
\label{l:sc}
\cite{hem-hem-hem:ctoappear:query-order-ph}
If ${\cal C}_1$ and ${\cal C}_2$ are classes such that
${\cal C}_1$ is
closed downwards under 
$\leq_m^{p,{\cal C}_2[1]}$
then 
$$\pconectwo = \pctwocone =\p^{({\cal C}_1,{\cal C}_2)}.$$
\end{lemma}

Now we are prepared to state and prove 
the
main theorem of this section,
Theorem~\ref{t:nxgeneral}.  It will follow easily from
this theorem that
$\rsnnp$ is equivalent to ordered access to NP and $\np \cap \conp$,
and also to parallel access to NP and $\npinterconp$---with one query
to each of NP and $\npinterconp$ allowed in each case.
Theorem~\ref{t:nxgeneral}'s proof uses the following technique.  The
theorem deals with $\rsncalc$, i.e., with a certain type of 1-truth-table
reduction.  A 1-truth-table has two bits of information---what to do
(accept versus reject) if the answer is yes, and what to do (accept
versus reject) if the answer is no---contained in the truth-table
itself.  Additionally, 
in $\rsncalc$ there is information in the (yes/no) answer from the
$\calc$ query.  The key trick in the proof (this occurs in the proof that
$\rsncalc \subseteq \funnyttcalc$) is to restructure this so that the effect
of the 1-truth-table reduction to a $\calc$ query is simulated by
one query each to $\np \cap \conp$ and $\calc$.  In effect, the $\np
\cap \conp$ returns, in its one-bit answer, enough information about
the two-bit truth-table that the base machine, working hand-in-hand
with the $\calc$ query, can make do with the one bit rather than two.

\begin{theorem}
\label{t:nxgeneral}
For every class \calc\ that
is closed 
downwards under $\leq_m^{p,\npinterconp [1]}$ we have
$${\rm R}_{1\hbox{-}{tt}}^{\cal SN}(\calc)=\p^{\npinterconp : \calc}
=\p^{\calc : \npinterconp}=\p^{(\npinterconp , \calc)}.$$
\end{theorem}

\proof
Note that the rightmost two equalities follow immediately from 
Lemma~\ref{l:sc}.

It remains to show that $\rsncalc=\funnycalc$.
Let us first show the inclusion from right to left. 
So suppose $L \in \funnycalc$,
as witnessed by DPTM $M$, $A \in \np \cap \conp$, and 
$B \in \calc$.
Without loss of generality, assume that $M$
always makes exactly one query to 
each of its oracles.
According to the definition of \snonettred\ reductions as given in 
Section~\ref{s:prelim}, we have to find a set $C \in \calc$ and a function 
$f \in \npsvt$ computing a string $y$~($=y(x)$) 
and a predicate $\alpha$~($=\alpha(x)$) such that
$$x \in L \iff \alpha(\chi_C(y)).$$ 
We specify $C$ by setting $C$ to equal $B$.
Let $y$ be the query that is asked
by $M^{A:B}(x)$ to $B$ when the first query is
answered correctly, and let 
$\alpha$ be a predicate defined by
$$\alpha(i) \iff 
\left( i=0 \land x \in L(M^{A:\emptyset})\right) \lor
\left( i=1 \land x \in L(M^{A:\sigmastar})\right).$$
Set $f(x)=\langle y,\alpha \rangle$ and note that $f \in \npsvt$ and also 
$x \in L \iff \alpha(\chi_B(y)).$ 
This proves $\funnycalc \subseteq \rsncalc$.

The proof will be completed
if we can
prove $\rsncalc \subseteq 
\funnyttcalc$; let us do so.
Let $L\in \rsncalc$.  Let 
$B \in \calc$ and $\npsvt$ function $f$
witness
(in the sense of Definition~\ref{d:sn})
that $L\in \rsncalc$.  So say $f(x) = 
\langle \langle X_1(x),X_2(x) \rangle , z(x) \rangle$.
In particular, let 
$(X_1(x),X_2(x))$---$X_1(x), X_2(x) 
\in \{{\rm A,R}\}$ where A stands for accept and 
R for reject---denote the truth-table that is output as the first 
component of $f(x)$, i.e., $X_1(x)$ denotes
the behavior (accept or reject) that occurs if 
the answer to the query to $\cal C$ is ``no''
and $X_2(x)$ denotes
the behavior (accept or reject) that occurs if 
the answer to the query to $\cal C$ is ``yes.''

We define sets $E \in \np \cap \conp$ and $F \in 
{\rm R}_m^{p, \npinterconp[1]}(\calc)$, and describe 
a DPTM $\widehat{M}$ such that $L=L({\widehat{M}}^{(E,F)})$.
Let 
\begin{quote}
\noindent
$E=\{x \condition (X_1(x),X_2(x)) \neq \mbox{(A,R)}\}$
\end{quote}
and
\begin{quote}
\noindent
$F=\{x \condition  (X_1(x),X_2(x)) =\mbox{(A,A)} 
\lor 
\left( (X_1(x),X_2(x))=\mbox{(R,A)} \land^{\mbox{\hspace{1pt}}} 
z(x)\in B\right)
\lor
\left( (X_1(x),X_2(x))=\mbox{(A,R)} 
\land^{\mbox{\hspace{1pt}}} z(x)\in B\right)\}$.
\end{quote}
Note that $E \in \np \cap \conp$ and $F \in \calc$ by our 
hypothesis that $\calc \supseteq  
{\rm R}_m^{p, \npinterconp[1]}(\calc)$.
Furthermore, let ${\widehat{M}}^{(E,F)}$ on input $x$ query ``$x \in E$?'' and 
``$x \in F$?'' and accept if and only if either both queries are answered 
``yes'' or both are answered ``no.''

Note that $L=L({\widehat{M}}^{(E,F)})$---as can 
easily be seen by considering each of the 
four cases (R,R), (R,A), (A,R), and (A,A).  Thus 
$\rsncalc \subseteq \funnyttcalc$,
which completes the proof of the theorem.~\qed

For any classes $\calc_1$ and $\calc_2$, let $\calc_1 \ominus
\calc_2 =_{\rm def} \{L \condition (\exists A \in \calc_1)\allowbreak
(\exists B \in \calc_2)\allowbreak [L=A-B]\}$.
\begin{theorem}\label{t:where-better-hypothesis}
For every class \calc\ that is closed 
downwards under $\leq_m^{p,\npinterconp [1]}$, it holds that
$$\p^{\calc [1]} \subseteq {\rm R}_{1\hbox{-}{tt}}^{\cal SN}(\calc)
\subseteq \calc \ominus \calc.$$
\end{theorem}

\proof
Note that by Theorem~\ref{t:nxgeneral}
we have 
${\rm R}_{1\hbox{-}{tt}}^{\cal SN}(\calc) = \p^{(\npinterconp , \calc)}$.
$\p^{\calc [1]} \subseteq {\rm R}_{1\hbox{-}{tt}}^{\cal SN}(\calc)$ 
follows immediately.
For the second inclusion, namely 
${\rm R}_{1\hbox{-}{tt}}^{\cal SN}(\calc)\subseteq \calc 
\ominus \calc$, suppose 
$L \in {\rm R}_{1\hbox{-}{tt}}^{\cal SN}(\calc)$ and thus, by 
Theorem~\ref{t:nxgeneral},
$L \in \p^{(\npinterconp , \calc)}$. 
Let $L \in \funnyttcalc$ be
witnessed by DPTM $M$, $A \in \np \cap \conp$, and $B \in \calc$.
Without loss of generality assume that $M$ makes, on every input, 
exactly one 
query to $A$ and exactly one query to $B$.
We describe two sets $F_1$ and $F_2$, both from \calc, such that $L=F_1-F_2$.
Before we come to the actual definition of $F_1$ and $F_2$ we introduce some 
notations that
are similar to the ones used in the proof of Theorem~\ref{t:nxgeneral}.
Let $(X_1(x),X_2(x))$---$X_1(x), X_2(x) 
\in \{{\rm A,R}\}$, where A stands for accept and 
R for reject---denote the 1-variable
truth-table with respect to the  query to $B$
of $M^{(A,B)}(x)$ (given that the first query is answered correctly).
That is,
\begin{description}
\item $X_1(x)$ is the outcome of $M^{(A,\emptyset)}(x)$, and
\item $X_2(x)$ is the outcome of $M^{(A,\sigmastar)}(x)$.
\end{description} 

Let $q_2(x)$ denote the query asked to $B$ by
$M^{(A,B)}(x)$.
\begin{quote}
\noindent
$F_1=\{x \condition (X_1(x),X_2(x))=\mbox{(A,A)}
\lor 
(X_1(x),X_2(x))=\mbox{(A,R)}
\lor  \left(  (X_1(x),X_2(x))=\mbox{(R,A)} \land 
q_2(x) \in B\right)\}$
\end{quote}
and
\begin{quote}
\noindent
$F_2=\{x \condition (X_1(x),X_2(x))=\mbox{(A,R)} \land q_2(x)\in B\}$.
\end{quote}

Note that both $F_1$ and $F_2$ are 
indeed in \calc, since both sets are clearly 
in ${\rm R}_m^{p,\np \cap \conp[1]}(\calc)$ and we know by assumption that 
this class equals \calc.
On the other hand one can easily verify that $L=F_1-F_2$ and thus 
$L \in \calc \ominus \calc$.~\qed

Let us apply to $\rsnnp$ the results just 
obtained.  The following well-known fact will be helpful.

\begin{lemma}\label{l:help}
${\rm R}_m^{p,\np \cap \conp[1]}(\np)=\np$.
\end{lemma}

\proof
The inclusion from right to left is clear.
Consider the inverse inclusion and note that clearly 
${\rm R}_m^{p,\np \cap \conp[1]}(\np) \subseteq {\np}^{\np \cap \conp[1]}$.
However, $\np  = \np^{\npinterconp}$ 
(first obtained
in~\cite{lon:j:nondeterministic-reducibilities}---which 
never states this but does subtly use it)
and thus   
${\rm R}_m^{p,\np \cap \conp[1]}(\np)=\np$.~\qed

From Lemma~\ref{l:help} and 
Theorems~\ref{t:nxgeneral} and~\ref{t:where-better-hypothesis}
we have the following two corollaries for $\rsnnp$.

\begin{corollary}
\label{c:nxspecific}
$\rsnnp = \funny =\funnyrev =\funnytt$.
\end{corollary}

\begin{corollary}\label{c:where}
$\pnpone \subseteq \rsnnp \subseteq {\rm DP}.$
\end{corollary}
 
Since $\rsnnp$ is closed under complementation
but DP is suspected not to be, the second 
inclusion probably is strict (we note in passing
that, due to the closure under complementation 
of $\rsnnp$, 
$\rsnnp \subseteq {\rm DP} \iff
\rsnnp \subseteq {\rm DP} \cap {\rm coDP}$).

\begin{corollary}\label{c:cons}  
If $\rsnnp={\rm DP}$ 
(equivalently, if 
$\rsnnp={\rm DP} \cap {\rm coDP}$) 
then the boolean hierarchy
collapses (and thus,
by~\cite{kad:joutdatedbychangkadin:bh},
the 
polynomial
hierarchy also collapses).
\end{corollary}

Though Corollary~\ref{c:cons} gives strong evidence that the second 
inclusion of Corollary~\ref{c:where} is strict, 
we know of no class collapse that follows from the assumption that the 
first inclusion is not strict  (though it is easy to directly construct 
an oracle relative to which the first 
inclusion is strict, and clearly the first inclusion must be strict in the 
relativized world we are going to construct in Section~\ref{s:completeness} 
in which $\rsnnp$ lacks m-complete sets).  
Can one prove that $\pnpone=\rsnnp$ implies some surprising collapse of 
complexity classes?

What types of sets are in \rsnnp?
Define ${\rm PrimeSAT}=\{\langle i,f \rangle \condition i \in {\rm PRIMES} 
\iff f \in {\rm SAT}\}$.
Clearly ${\rm PrimeSAT} \in \funnytt$ and 
thus, by Corollary~\ref{c:nxspecific}, ${\rm PrimeSAT} \in \rsnnp$.
On the other hand, 
${\rm PrimeSAT}\in \p^{({\rm ZPP} \cap \up \cap \coup,\np)}$ 
(since ${\rm PRIMES} \in {\rm ZPP} \cap \up \cap \coup$ 
\cite{adl-hua:cOUTbyBOOKwhichIsInCompanionBib:primes,fel-kob:c:self-witnessing}),
so it seems somewhat unlikely that PrimeSAT is m-complete for \rsnnp.
In fact, though 
(see the discussion in 
Section~\ref{s:completeness}) 
\rsnnp\ robustly has 
2-tt-complete sets, 
nonetheless $\rsnnp$ may well lack 
1-tt-complete sets.
In fact, we will in the next section 
construct a relativized world in which 
\rsnnp\ has no 1-tt-complete set.

Finally, we note that $\funny = \rsnnp$ is
a case where guarded database (oracle) access seems
more powerful than standard access.
So-called guarded reductions were introduced by 
Grollmann and 
Selman~\cite{gro-sel:j:complexity-measures} 
(there called ``smart reductions''), and
were further investigated 
by Cai,
Hemaspaandra (then Hemachandra),
and Vysko\v{c}~\cite{cai-hem-vys:b:promise}.
In light of Corollary~\ref{c:nxspecific} we will now
look at guarded oracle access to $\np \cap \conp$ in
the context of 
Definition~\ref{d:new}, in order to see whether guarded 
access yields yet another equivalent form 
of $\rsnnp$.  We will see that it seems not 
to.  

Let \smartfunnyrev\ be the class of languages that are recognized 
by some DPTM $M$ that 
makes two sequential queries, 
the first to some \np\ set $A$, and the 
second to some NP set $B$, and such that 
it also holds that there is another NP set $C$
such that, for all $y$:
\begin{quote}
\noindent
if there is an $x$ such
that $M^{A:B}(x)$ given the correct answer to its query to $A$ (if 
any) queries ``$y\in B$?''~then:  $y \in B 
\iff y \not\in C$.
\end{quote}
In other words, $M$ is allowed an ordinary 
query to NP, followed by a query that must 
be ``$\npinterconp$-like.''  However, this 
is not necessarily the same as allowing an
$\npinterconp$ query (indeed, see Theorem~\ref{t:smart}).
The key point is that on strings never asked by $M^{A:B}$ to $B$, $B$ and
$C$ need not be complementary.  Informally, $M$ and $A$ guard $B$
against queries where $B$ might fail to complement 
$C$.\footnote{\protect\singlespacing{}There
is no need to similarly define \smartfunny\ and \smartfunnytt,
as it is clear that 
$\rsnnp = \smartfunny = \smartfunnytt$.  This is just 
a reflection of the known fact that, in most contexts,
if the ``first'' query is guarded it can as well be unguarded
as the very fact that it is asked is a certificate that the 
query obeys the appropriate promise~\protect\cite{cai-hem-vys:b:promise}.}

\begin{theorem}\label{t:smart}
$\rsnnp = \funnyrev \subseteq \dpintercodp \subseteq 
\smartfunnyrev$.
\end{theorem}

\proof
The first equality is part of Corollary~\ref{c:nxspecific}.
The inclusion $\rsnnp \subseteq \dpintercodp$ follows immediately from 
Corollary~\ref{c:where} as noted after that corollary.

It remains to show $\dpintercodp \subseteq \smartfunnyrev$. 
Let $L \in \dpintercodp$. 
So there exist NP sets $L_1$, $L_2$, $L_3$, and $L_4$ such that 
$L=L_1-L_2=\overline{L_3-L_4}$.
Following
Cai et 
al.~\cite{cai-gun-har-hem-sew-wag-wec:j:bh1}, we can without loss of 
generality assume that $L_2 \subseteq L_1$ and $L_4 \subseteq L_3$. 

We make the following observations: 
\begin{enumerate}
\item 
$(x \notin L_2 \land x \notin L_4) \Rightarrow
(x \in L_1 \iff x \notin L_3)$, and
\item $( x \in L_2 \lor x \in L_4) \Rightarrow 
(x \in L_4 \iff x \notin L_2)$.
\end{enumerate}

Define 
$E=\{x \condition x \in L_2 \lor x \in L_4\}$ and
$F=\{\langle x,0 \rangle \condition x \in L_1\} \cup 
\{\langle x,1 \rangle \condition x \in L_4\}$,
and note that clearly $E$ and $F$ are in \np.

Let $M$ be a DPTM that on input $x$ first queries ``$x \in E$?''\ and if it 
gets the answer $i$, where $i \in \{0,1\}$ and 0 stands for ``no'' and 1 
stands for ``yes,'' queries ``$\langle x,i \rangle \in F$?''\ and accepts if 
and only if this second query is answered ``yes.''

By our above observations we know that $M$ ``smartly'' accesses $F$ and hence 
$L \in \smartfunnyrev$.~\qed

\section{Completeness}\label{s:completeness}

In this section we prove that there is a relativized world in which 
\rsnnp\ has no 1-tt-complete sets (and thus no 
m-complete sets).  We will note that \rsnnp\ robustly
has 2-tt-complete sets.  
Thus we show even more, namely that \rsnnp\ distinguishes robust 
1-tt-completeness from robust 2-tt-completeness (and thus it
also distinguishes robust m-completeness from robust T-completeness).

To discuss relativized completeness we must define relativized
reductions and the natural relativizations of our classes.  So that
our theorems are fair, we choose full relativizations
(see~\cite{rog:b:rft}), i.e., relativizations in which both the
reductions and the classes may access the oracle.  However, as
Theorem~\ref{t:all-same} will show, many different statements
regarding completeness---some involving partial relativizations---are
equivalent.  In fact, we will make use of some of these equivalences
in proving our result.

\begin{definition} \label{d:self}
\begin{enumerate}
\item \label{p:one}  Let $\snaonettred$ be as in 
Definition~\ref{d:sn}, except with $\npsvt$
replaced by $\npsvta$.
\item (Full relativization of \rsnnp)\ $\rsnnpa =_{\rm def}
{{\rm R}_{1\hbox{-}{tt}}^{{\cal SN},A}(\npa)}$, i.e.,
$\{L \condition (\exists C \in \npa)\allowbreak
[L \snaonettred C]\}$.
\item Let $\onettreda$ (respectively, 
$\onettred$~\cite{lad-lyn-sel:j:com}) 
be as in Part~\ref{p:one} of the present
definition (respectively, as in Definition~\ref{d:sn})
except with $\npsvta$ (respectively, $\npsvt$) 
replaced by $\fpta$ (respectively, $\fpt$), where $\fpt$
denotes the deterministic polynomial-time computable
functions.\footnote{\protect\singlespacing{}This definition is equivalent
to the more traditional
``generator plus evaluator'' 1-truth-table definition with,
as is natural,
both the generator and the evaluator relativized.}
Many-one and 2-truth-table reductions are relativized 
in the obvious analogous ways.
\item (Full relativization of $\funnytt$)\ $\funnytta =_{\rm def}
\{L \condition$ $L$ is recognized by a deterministic polynomial-time 
Turing machine that 
makes, in parallel,
at most one query to $\np^A \cap \conp^A$ and 
at most one query to 
$\np^A$, and that additionally has---before the 
parallel round or after the parallel round or
both---unlimited access to $A\}$.
\end{enumerate}
\end{definition}

\begin{theorem}\label{t:all-same}  Let $A$ be any set.
All of the following twelve statements are equivalent.
\begin{enumerate}
\item \label{p:list-one} $\rsnnpa$ has $\onettreda$-complete sets.
\item $\rsnnpa$ has $\onettred$-complete sets.
\item $\rsnnpa$ has $\manyonea$-complete sets.
\item \label{p:list-four} $\rsnnpa$ has $\manyone$-complete sets.
\item[5--8.]  The same as Parts~\ref{p:list-one}--\ref{p:list-four},
with $\rsnnpa$ replaced by $\funnytta$.
\item[9--12.]  The same as Parts~\ref{p:list-one}--\ref{p:list-four},
with $\rsnnpa$ replaced by $\partialfuna$.
\end{enumerate}
\end{theorem}

Theorem~\ref{t:all-same} follows from Lemma~\ref{l:three-same},
Lemma~\ref{l:m-1tt}, and the fact that 
$\rsnnpa$ is clearly closed downwards under $\onettreda$ reductions.

\begin{lemma}\label{l:three-same}
For each set $A$, $\rsnnpa = \funnytta = \partialfuna$.
\end{lemma}

Lemma~\ref{l:three-same} is essentially 
a relativized version of part of 
Corollary~\ref{c:nxspecific}, plus the 
observation that the technique used in the second half of 
the proof of that result (that is, the second 
half of the proof of Theorem~\ref{t:nxgeneral}, in
the case $\calc = \np$) in fact can easily show 
not just $\rsnnpa \subseteq \funnytta$, but even
$\rsnnpa \subseteq \partialfuna$.

Most non-completeness proofs of the 
Sipser/Regan/Borchert school (i.e., proofs based on tainting
enumerations) use a bridge between the existence of
$\calca$-$\manyonea$-complete sets and the existence of
$\calca$-$\manyone$-complete sets. Here, we 
extend that link to also embrace 1-truth-table reductions.

\begin{lemma}\label{l:m-1tt}
Let \cald\ be any class (quite possibly a relativized
class, such as $\rsnnpa$) that is 
closed downwards under \onettreda\ reductions.
Then the following four statements are 
equivalent: (a)~\cald\ has $\manyone$-complete sets,
(b)~\cald\ has 
$\manyonea$-complete sets,
(c)~\cald\ has 
$\onettred$-complete sets,
(d)~\cald\ has 
$\onettreda$-complete sets.
\end{lemma}

\proof
Since every $\manyone$-complete set for \cald\ is also 
$\manyonea$-, $\onettred$-, and $\onettreda$-complete for \cald, 
we have to show only that if \cald\ is closed 
downwards under \onettreda\ reductions and \cald\ has 
$\onettreda$-complete sets, then \cald\ also 
has $\manyone$-complete sets. 

So let $L$ be $\onettreda$-complete for \cald. 
Define
$L'=\{\langle x,i,0^{k}\rangle \condition M_i^{L,A}(x)$, when run 
allowing at most one oracle query to $L$ during the
run but allowed unlimited access to $A$,  
accepts within $k$ steps$\}$.
Note that $L' \onettreda L$, and thus 
$L' \in \cald$.

Also, $L'$ is \manyonered-hard for \cald.
To verify this let $B \in \cald$.
Then by the $\onettreda$-completeness of $L$, $B \onettreda L$, say via 
machine $M_j$. 
So $f(x)=\langle x,j,0^{|x|^j+j}\rangle$ is a \manyonered\ reduction from 
$B$ to $L'$.
This proves that $L'$ is $\manyone$-complete for \cald.~\qed

As is standard in the Sipser/Regan/Borchert approach to 
establishing non-completeness, we wish to characterize
the existence of complete sets via the issue of the existence of a 
certain index set.  Lemma~\ref{l:key-enum} does this.
Since it is quite similar to the analogous 
lemmas in 
previous non-completeness papers 
(see, e.g., \cite[Lemmas~2.7 and~4.2]{har-hem:j:up})
we do not include the proof.  We do, however, mention
the following points.  The lemma draws freely on
Theorem~\ref{t:all-same}.  Also, the claim in Lemma~\ref{l:key-enum}
regarding $\p$ and $\p^A$ being equivalent (in that context)
is an invocation of a trick from
the literature~\cite[p.~134]{har-hem:j:up}.

\begin{lemma}\label{l:key-enum}
For every oracle $A$, \rsnnpa\ has 
\onettreda-complete sets if and only if there 
exists a \p\ set (equivalently, a $\p^A$ set) $I$ of index quadruples such that
\begin{enumerate}
\item \label{a}$(\forall \quadruple)[\quadruple \in I \Rightarrow 
L(N_j^A)=\overline{L(N_k^A)}]$, and
\item \label{b}$\partialfuna
=\{L\left( M_i^{(L(N_j^A),L(N_l^A))}\right) \condition 
(\exists k)\allowbreak[\quadruple \in I]\}$, and
\item \label{c} $(\forall \quadruple)[ \quadruple \in I \Rightarrow
(\forall x)[$In the run of $M_i^{(L(N_j^A),L(N_l^A))}(x)$, 
$M_i$ makes at most one 
round of truth-table queries and that round consists of at most 
one query (simultaneously) to each part of its
``$\left( L(N_j^A),L(N_l^A)\right)$'' oracle$]]$.
\end{enumerate}
\end{lemma}

We now prove our non-completeness claim.

\begin{theorem}
\label{t:no-complete}
There is a recursive oracle $A$ such 
that \rsnnpa\ lacks
\onettreda-complete sets.
\end{theorem}

\proof
The proof consists of the construction of a 
recursive set $A$ such that 
there exists no \p\ set $I$ having properties~\ref{a},~\ref{b},
and~\ref{c} of Lemma~\ref{l:key-enum}.

Let $\{\widehat{M}_i\}$ be a standard 
enumeration of deterministic polynomial-time Turing
machines.
Let $\{N_i\}$ be the enumeration of
nondeterministic polynomial-time oracle Turing machines described in 
Section~\ref{s:prelim}. 
Let $\{M_i\}$ be an enumeration of
deterministic polynomial-time oracle Turing machines 
satisfying all the properties of the enumeration
$\{M_i\}$ described in
Section~\ref{s:prelim} and having the additional
property that every 
machine from $\{M_i\}$, on every input $x$, 
makes exactly one parallel round of 
exactly one query to each of its two oracles.

It is clear that there is such an enumeration 
$\{M_i\}$ and 
that 
this enumeration ensures that for each oracle $A$, 
for each $i$, $j$, and $l$, and for
each input $x$, the queries asked by 
$M_i^{(L(N_j^A),L(N_l^A))}(x)$ to $L(N_j^A)$ and $L(N_l^A)$ are 
independent of $A$.

To show that there exists no \p\ set $I$ having 
properties~\ref{a},~\ref{b}, and~\ref{c} of Lemma~\ref{l:key-enum}, 
we will diagonalize against all \p\ machines in such a way that 
eventually for every \p\ machine $\widehat{M}_h$ (let $h_e$ denote the $e$th 
string, $h_e=\quadruple$, accepted by $\widehat{M}_h$) at least one of the 
following holds:
\begin{description}
\item[Goal 1]  $\widehat{M}_h$ accepts some 
string $h_{\hat{e}}=\quadruplehats$ 
such 
that $L(N_{\widehat{j}}^A) \not= \overline{L(N_{\widehat{k}}^A)}$. 
In this case, 
$\widehat{M}_h$ does not accept a set of index quadruples $I$ having 
property~\ref{a} of Lemma~\ref{l:key-enum}.
\item[Goal 2] 
There exists a set 
$D_h \in \partialfuna$ 
such that, 
for every $h_e=\quadruple$ accepted by $\widehat{M}_h$, 
$D_h \not= L\left(  M_i^{(L(N_j^A),L(N_l^A))}\right)$. 
($D_h$ will be explicitly defined later in the proof.)
Thus $\widehat{M}_h$ does  not accept a set of index quadruples covering 
$\partialfuna$ and hence $\widehat{M}_h$ accepts 
a set not having property~\ref{b} of 
Lemma~\ref{l:key-enum}.
\end{description}

In the proof we will build a list, $\canmac$, of canceled pairs 
$(h,e)$. We add a pair $(h,e)$, 
$h_e=\quadruple$, to the list $\canmac$ when 
either Goal~1 has been met for $h$ (in this 
case all $(h,e')$, $e'\in\naturalnumber$, are 
marked canceled), or $h_e$ is 
consistent with Goal~2 for $h$
(i.e., with $h_e = \quadruple$, 
$D_h \not= L\left(  M_i^{(L(N_j^A),L(N_l^A))}\right)$).  
We describe the former case as a type 1 cancellation
and the latter case as a type~2 cancellation.

Now let us define the languages $D_h$.
For every $h \geq 1$, let
\begin{eqnarray*}
D_h =  \{ & 0^n \condition  &
n \geq 3 
\land \\ & & 
(\exists k \geq 1)[n=(p_h)^k] \land \\ & & 
[((\exists y)[|y|=n-2 \land 00y \in A]) \iff 
((\exists z)[|z|=n-2 \land 11z \in A])]  \}, 
\end{eqnarray*}
where $p_h$ denotes the $h$th prime.

We will always construct $A$ so that, for all $h$
such 
that for no $\widehat{e}$ is $(h,\widehat{e})$ ever 
involved in 
a type~1 cancellation, it holds that: 
for each $n$ such that, for some $k$, $n=(p_h)^k$:
$${\bf \circledast} \qquad ((\exists y)[|y|=n-2 \land 00y \in A]) \iff 
\neg((\exists y)[|y|=n-2 \land 01y \in A]).$$
Note 
that $\circledast$ will ensure that---for those $h$ that are 
never involved in a type~1 cancellation---each such
$D_h$ will belong to $\partialfuna$
(namely, as the right-hand side of the ``$\Longleftrightarrow$''
of the definition of
$D_h$ is an $\npa$ type query, and the left-hand side of the
``$\Longleftrightarrow$'' of 
the definition of $D_h$ is in effect made into an $\npa \cap
\conpa$ type query by $\circledast$).

{\em Construction of 
{$A =_{\rm def} \bigcup_{m\geq 0} A_m$}:}

\noindent\underline{Stage 0:} $\canmac=\emptyset$, $A_0=\emptyset$

\noindent\underline{Stage $m$, $m > 0$:} 
Consider the uncanceled pair $(h,e)$, $e<m$, $(h,e) \notin \canmac$, 
for which $h+e$ is smallest.

\noindent{\bf If} 
\begin{enumerate}
\item[(i)] no such pair exists, or
\item[(ii)] any length $m$ string was queried at any previous 
stage,\footnote{\protect\singlespacing{}This condition
is {\em never\/} satisfied.  We include it just to to emphasize
that it is not an issue.  The reason it is not an issue 
is discussed in Footnote~\protect\ref{f:magic-coding}.}
or
\item[(iii)] $m$ is not of the form $(p_h)^q$ for some 
$q \in \naturalnumber$, where $p_h$ is the $h$th prime, or
\item[(iv)] the pair $(h,e)$, $h_e=\quadruple$, that the above rule
chooses (if any) is 
such that $3\left(  (m^i+i)^{\max(j,k,l)} + 
\max(j,k,l)\right) \geq 2^{m - 16}$,\footnote{\protect\singlespacing{}If 
$3\left(  (m^i+i)^{\max(j,k,l)} + 
\max(j,k,l)\right) < 2^{m - 16}$ it (easily) holds that 
on the run of $M_i^{(L(N_j^A),L(N_l^A))}(0^m)$
the maximum number of queries to $A$ made on any one path
of $N_j^A(q_1)$ plus
the maximum number of queries to $A$ made on any one path
of $N_k^A(q_1)$ plus
the maximum number of queries to $A$ made on any one path
of $N_l^A(q_2)$ is less than $2^{m-16}$, where $q_1$ 
and $q_2$ respectively denote the queries to 
$L(N_j^A)$ and $L(N_l^A)$ made by 
$M_i^{(L(N_j^A),L(N_l^A))}(0^m)$.}
\end{enumerate}

\noindent {\bf then} set $A_m=A_{m-1}\cup \{0^m\}$ 
(in order to maintain 
$\circledast$)\footnote{\protect\singlespacing{}\protect\label{f:magic-coding}Note 
that 
$0^m$ is free to put in, and so we do not have 
to search for 
some unqueried string chosen
from $00\Sigma^{m-2}$.  The reason we can use 
$0^m$ is that, as will soon become clear, each time 
a stage $m'$ touches strings of length greater than
$m'$, that stage then ``jumps forward in time'' 
(while maintaining appropriate codings in light
of $\circledast$) to 
a stage $m''$ such that $m''$ is strictly greater 
than the length of any string queried at stage $m'$. 
In short, at the current point in the proof, {\em no\/}
strings of length $m$ have been queried or frozen, and thus
in particular $0^m$ has never been queried or frozen.}
and go to Stage $m+1$.

Otherwise (i.e., if the ``if'' above is not 
satisfied),
let $(h,e)$ be the selected pair, $h_e=\quadruple$.
Define $\gamma = (m^i+i)^{\max(j,k,l)} + 
\max(j,k,l)$.  Note that $\gamma$ is an upper bound
on the length of the strings in $A$ that can be queried
at this stage by any of $M_i$, $N_j$, $N_k$ (when run 
on whatever query $N_j$ is run on),
and $N_l$.  Let $\protm$ denote the class of all sets $E$
having exactly one string at each length $\widehat{i}$,
$m<\widehat{i}\leq \gamma$, and such that $E$ meets $\circledast$
for each length $\widehat{i}$,
$m<\widehat{i}\leq\gamma$, and such that $E$ has no strings
other than those just described.  As noted in
Footnote~\ref{f:magic-coding}, no strings at lengths in
this range have yet been queried, so we do not have 
to worry about already-frozen strings existing at these 
lengths.  The reason we must include $\protm$ in
the condition that distinguishes between Case~1
and Case~2 is that one of the two possible ways
the ``$\neq$'' of the Case~1 test can be satisfied (namely,
if there is a string that is not in the set on the left-hand
side and that is in the set on the right-hand side) requires
us to freeze {\em rejecting\/} behavior of $N_j$ and $N_k$.
This involves freezing (i.e., fixing 
permanently the membership
status regarding the oracle)
an exponential number of strings---enough
to ruin any attempts to satisfy $\circledast$ at length $m+1$ 
for example.  Looking at $\protm$ allows us to handle $m$ and
the problem $m$ might cause at lengths $m+1,\,m+2,\,\cdots,\,\gamma$
in an integrated fashion.  In particular, simultaneously
with choosing a length~$m$ extension we will choose an appropriate
extension that handles the $\circledast$ coding for lengths
$m+1,\,m+2,\,\cdots,\,\gamma$.
There are two cases.

\begin{description}
\item[Case 1:]
For some $B \subseteq \Sigma^m$, for some $B'\in \protm$,
\begin{center}
$\left(L\left( N_j^{A_{m-1} \cup B \fixupa}\right)
\right)^{\leq m^i+i} \not= 
\left(\overline{L\left( N_k^{A_{m-1} \cup B \fixupa}\right)}
\right)^{\leq m^i+i}$.\footnote{\protect\singlespacing{}The $m^i+i$ bounds
here are to ensure that the construction yields a 
recursive oracle.}
\end{center}

\noindent 
In this case, we have a type~1 
cancellation.  
So, for each $e' \in \naturalnumber$, add $(h,e')$ to $\canmac$.
For each $m \leq p \leq \gamma$,
set $A_p=A_{m-1} \cup B \fixupa$.  Go to stage $\gamma + 1$.
Note that in this case we do not 
necessarily maintain $\circledast$ at the current length in the 
construction of $A$.
However,
since we cancel ``the entire machine $M_h$'' (by canceling all pairs 
$(h,e')$) and thus have successfully diagonalized against machine $M_h$,
achieving Goal 1, we do not need in this case to 
argue regarding the set $D_h$ and thus do not have to ensure that for this 
specific $h$, $D_h \in \partialfuna$.  We have, however, 
maintained $\circledast$ at lengths $m+1,\,m+2,\,\cdots,\,\gamma$.

\item[Case 2:]
For every $B \subseteq \Sigma^m$, for every $B' \in \protm$,
\begin{center}
$\left(L\left( N_j^{A_{m-1} \cup B \fixupa
}\right)
\right)^{\leq m^i+i} = 
\left(\overline{L\left( N_k^{A_{m-1} \cup B \fixupa
}\right)}
\right)^{\leq m^i+i}$.
\end{center}

Note that this tells us that with respect to all oracles of the form 
$A_{m-1} \cup B \fixupa$, $B \subseteq \Sigma^m$
and $B'\in\protm$, $N_j$ and $N_k$ are complementary
at all lengths that might be queried by 
$M_i^{(L(N_j^A),L(N_l^A))}(0^m)$.
We are now going to exploit this fact in order to achieve a type~2
cancellation of the pair $(h,e)$.  Note that we now must be careful 
in adding strings to our oracle set, since we have to maintain $\circledast$
at length $m$.  To satisfy $\circledast$ at 
lengths $m+1,\,m+2,\,\cdots,\,\gamma$, let
$B''$ be any fixed member of $\protm$;  we will
make $B''$ a part of the oracle extension.

Recall that the machine $M_i$ makes one query round consisting of 
one query each (in parallel)  
to $L(N_j^A)$ and $L(N_l^A)$.
We consider the action of
$$M_i^{(L(N_j^{A_{m-1}\fixupz}),L(N_l^{A_{m-1}\fixupz}))}(0^m).$$
There are four cases depending on the answers to the two queries 
$M_i^{(L(N_j^{A_{m-1}\fixupz}),L(N_l^{A_{m-1}\fixupz}))}(0^m)$ makes.
We henceforward assume that 
the query to $L(M_j^{A_{m-1}\fixupz})$ is answered 
``yes.''
The other case (``no'') is omitted as it is 
very analogous (note that in 
the case we are in, i.e., Case 2, if the answer to the query to 
$L(N_j^{A_{m-1}\fixupz})$ is ``no'' then 
the same query to $L(N_k^{A_{m-1}\fixupz})$ is 
answered ``yes'').

Regarding the query to $L(N_l^{A_{m-1}\fixupz})$, the hard case 
is if it gets the 
answer ``no,'' as if it gets the answer ``yes,'' we freeze the 
lexicographically first accepting path (call it $\varsigma$) 
and then perform a simpler version of the proof that is about to 
come.\footnote{\protect\singlespacing{}Essentially 
one has to change the 
definition of $guarded_m$ 
(in order to include the queries made to $A$ along $\varsigma$) and do
Case 2a with slight modifications.}
So let us focus just on the case where 
$L(N_l^{A_{m-1}\fixupz})$ says ``no.''

Freeze the lexicographically first accepting path $\varrho$ of 
$N_j^{A_{m-1} \fixupz}$ on the input on which it is called.
Let $guarded_m$ be
all queries to $A$ made on path $\varrho$.

\begin{description}
\item[Case 2a:] There is some string, $\alpha$, in 
$(00\Sigma^{m-2} \cup 01\Sigma^{m-2}) - guarded_m$
such that $N_l^{A_{m-1} \fixupz
\cup \{\alpha\}}$ accepts 
when run on 
the query asked to it by 
$M_i^{(L(N_j^{A_{m-1}  \fixupz \cup \{\alpha \}}),
L(N_l^{A_{m-1}  \fixupz \cup \{\alpha \}}))}(0^m)$.

\noindent {\bf If} either

\hfil
\begin{minipage}{12cm}
(a) $M_i^{(L(N_j^{A_{m-1} \fixupz \cup \{\alpha \}}),
L(N_l^{A_{m-1} \fixupz \cup \{\alpha \}}))}(0^m)$
accepts and the first two bits of $\alpha$ are~00, or \\
(b) $M_i^{(L(N_j^{A_{m-1} \fixupz \cup \{\alpha \}}),
L(N_l^{A_{m-1} \fixupz \cup \{\alpha \}}))}(0^m)$
rejects and the first two bits of $\alpha$ are~01,
\end{minipage}

{\bf then} we have achieved a type~2 
cancellation for $(h,e)$, i.e.,
$h_e$ codes a machine that does not accept $D_h$ (actually,
$D_h$ defined with respect to $A_m$ as we are about to 
define $A_m$, but the rest of the construction of $A$ will
not alter the achieved cancellation).
For each $m \leq p \leq \gamma$, 
set $A_p=A_{m-1} \fixupz \cup \{\alpha \}$ and add $(h,e)$ to $\canmac$.
Note that we have maintained $\circledast$ at the current length,
and at lengths $m+1,\cdots,\gamma$.
Go to stage $\gamma+1$.

Otherwise (i.e., if the ``if'' above is not 
satisfied) find a string $z$, $z \in 11\Sigma^{m-2}$,
such that $z \notin guarded_m$ and $z$ is not queried on the 
lexicographically first accepting path of 
$N_l^{A_{m-1} \fixupz \cup \{\alpha\}}(q')$, 
where $q'$ is 
the query asked to 
$N_l^{A_{m-1} \fixupz \cup \{\alpha\}}$
by 
$M_i^{(L(N_j^{A_{m-1}  \fixupz \cup \{\alpha \}}),
L(N_l^{A_{m-1}  \fixupz \cup \{\alpha \}}))}(0^m)$. 
Note that such a string $z$ easily exists by (iv) 
(from beginning of stage $m$).

Note that the following two items hold:
\begin{enumerate}
\item Either
\hfill   
\begin{minipage}[t]{11cm}
(a) $M_i^{(L(N_j^{A_{m-1} \fixupz \cup \{\alpha \}}),
L(N_l^{A_{m-1}\fixupz \cup \{\alpha \}}))}(0^m)$
accepts and the first two bits of $\alpha$ are 01, or \\
(b) $M_i^{(L(N_j^{A_{m-1}  \fixupz \cup \{\alpha \}}),
L(N_l^{A_{m-1} \fixupz  \cup \{\alpha \}}))}(0^m)$
rejects and the first two bits of $\alpha$ are 00.
\end{minipage}
\item $z$ is 
not queried on the lexicographically first accepting path of 
$N_j^{A_{m-1}  \fixupz }$ 
when 
run on 
the query asked to it by 
$M_i^{(L(N_j^{A_{m-1}  \fixupz \cup \{\alpha \}}),
L(N_l^{A_{m-1}  \fixupz \cup \{\alpha \}}))}(0^m)$, 
and is not
queried on the lexicographically first accepting path of $N_l^A$
when run on the query asked by
$M_i^{(L(N_j^{A_{m-1}  \fixupz \cup \{\alpha \}}),
L(N_l^{A_{m-1}  \fixupz \cup \{\alpha \}}))}(0^m)$.
\end{enumerate}

For each $m\leq p \leq \gamma$,
set $A_p=A_{m-1}  \fixupz \cup \{\alpha \} \cup \{z\}$. 
We have achieved a type~2
cancellation of $(h,e)$, so we add 
$(h,e)$ to $\canmac$.
Note that we have maintained $\circledast$ at  lengths
$m,\,m+1,\,\cdots,\,\gamma$.
Go to stage $\gamma+1$.

\item[Case 2b:] For each string $\alpha$ in 
$(00\Sigma^{m-2} \cup 01\Sigma^{m-2}) - guarded_m$
it holds that $N_l^{A_{m-1}  \fixupz 
\cup \{\alpha\}}$ rejects 
when run on 
the query asked to it by 
$M_i^{(L(N_j^{A_{m-1}  \fixupz \cup \{\alpha \}}),
L(N_l^{A_{m-1}  \fixupz \cup \{\alpha \}}))}(0^m)$.

If 
$M_i^{(L(N_j^{A_{m-1} \fixupz }),L(N_l^{A_{m-1} \fixupz }))}(0^m)$ 
accepts then 
let $\hat{w}$ be any string in $00\Sigma^{m-2}-guarded_m$ and 
if 
$M_i^{(L(N_j^{A_{m-1} \fixupz }),
L(N_l^{A_{m-1} \fixupz }))}(0^m)$ 
rejects then 
let $\hat{w}$ be any string in $01\Sigma^{m-2}-guarded_m$. 

By the same reason 
(i.e., item~(iv) from earlier in the proof) that
we can find a string $z$ in the ``otherwise'' part of Case~2a
we can find such a string $\hat{w}$.
For each $m\leq p \leq \gamma$, 
set $A_p=A_{m-1}  \fixupz 
\cup \{\hat{w}\}$ and note that we again have achieved 
a type~2 cancellation for $(h,e)$.
Add $(h,e)$ to $\canmac$.
Note that we have maintained $\circledast$ at the lengths
$m,\,m+1,\,\cdots,\,\gamma$.
Go to stage $\gamma+1$.
\end{description}
\end{description}

{\em This ends the construction of $A$}

Note that for
each $h_e = \quadruple$ it holds
that for all sufficiently large $m$ the ``$\geq$'' 
in~(iv) at the beginning of stage $m$ fails (as 
all polynomials in $m$ are $o(2^m)$). So
the above construction ensures that any given pair $(h,e)$ 
is eventually canceled, as once $h+e$ becomes the 
smallest among all uncanceled pairs it is clear
it will eventually be canceled.  Also,
for every $h$ we achieve either Goal 1 at some stage 
of the construction or we maintain $\circledast$ in the construction of $A$ at 
all lengths $m$ for 
which $m=(p_h)^k$ for some $k$ (where $p_h$ is the $h$th prime). 

The construction yields an oracle $A$ such 
that any polynomial-time machine $\widehat{M}_h$ either accepts at least one 
string $h_e=\quadruple$ such that $L(N_j^A) \not= L(N_k^A)$ 
(in this case $\widehat{M}_h$ does 
not accept a set of index quadruples having property~\ref{a} of 
Lemma~\ref{l:key-enum}) or else none of the machines (with corresponding 
oracles) whose quadruples are members
of the set $L(\widehat{M}_h)$ accept the language $D_h$. 
(Note that in this latter case, $D_h$ is in $\partialfuna$ since 
for every $h$ we maintain $\circledast$ in the construction of $A$ unless
we have a type~1 cancellation of $M_h$.)

This proves that no polynomial-time machine can accept a set of index 
quadruples $I$ having properties~\ref{a},~\ref{b},
and~\ref{c} of 
Lemma~\ref{l:key-enum}.
Thus, by Lemma~\ref{l:key-enum}, \rsnnpa\ has no 
$\onettreda$-complete sets.  Finally, by our 
construction, $A$ is clearly recursive (and, as is 
often the case regarding recursive oracles in 
complexity-theoretic constructions, one could make stronger
claims regarding its time complexity).~\qed

For each $A$, since $\rsnnpa \subseteq \rspecialtwoa$
and $\npa\subseteq \rsnnpa$, it 
follows that $\rsnnpa$ clearly has $\twottreda$-complete
sets.  In particular, all sets $\manyonea$-complete (or even
$\onettreda$-complete) for $\npa$ are $\twottreda$-complete
for $\rsnnpa$.  (We note in
passing that it is also immediately clear that, for each $A$,
$\rsnnpa$ has $\snaonettred$-complete
sets, e.g., all sets $\manyonea$-complete 
for $\npa$.)
We may now state the following from
Theorem~\ref{t:no-complete}.

\begin{corollary} There is a recursive oracle $A$
so that $\rsnnpa$ has $\twottreda$-complete sets
but has no
$\onettreda$-complete sets.
\end{corollary}

Weakening this, we have the following.

\begin{corollary} There is a recursive oracle $A$
so that $\rsnnpa$ has $\leq_T^{p,A}$-complete sets
but has no
$\manyonea$-complete sets.
\end{corollary}

In summary, 
we showed that \rsnnp\ distinguishes robust m-completeness from robust 
T-completeness.
Indeed it distinguishes robust 1-tt-completeness from robust 
2-tt-completeness. 
We conjecture that  
\dpintercodp\ 
will also 
distinguish robust 1-tt-completeness 
from 2-tt-completeness. 
However, note that this does not generalize to a claim that
${\rm BH}_k \cap {\rm coBH}_k$ 
(where ${\rm BH}_k$ is the $k$th level of the boolean hierarchy---see
\cite{cai-gun-har-hem-sew-wag-wec:j:bh1} for the 
definition of ${\rm BH}_k$)
distinguishes robust $k-1$-tt-completeness from 
robust $k$-tt-completeness.
In fact, it is clear that, for each $k \geq 2$, 
${\rm BH}_k \cap {\rm coBH}_k$ robustly has 2-tt-complete sets
(in fact, for $k\geq 2$, it 
follows from the structure of
the boolean hierarchy that all $\manyonea$-complete sets
for ${\rm BH}_{k-1}$ are $\twottreda$-complete for 
${\rm BH}_k \cap {\rm coBH}_k$).

{\samepage
\begin{center}
{\bf Acknowledgments}
\end{center}
\nopagebreak
\indent
We are deeply indebted to Gerd Wechsung for both 
his kind encouragement and his valuable guidance.
We thank Maren Hinrichs for enjoyable conversations,
Alan Selman for helping us with the history of 
strong nondeterministic reductions, and an anonymous referee for 
helpful comments.

}%

\singlespacing

\bibliographystyle{alpha}

\newcommand{\etalchar}[1]{$^{#1}$}

\end{document}